\documentclass[twocolumn,english,showpacs,superscriptaddress,prl]{revtex4-1}
\usepackage{hyperref}
\usepackage[latin9]{inputenc}
\usepackage{color}
\usepackage{float}
\usepackage{textcomp}
\usepackage{amstext}
\usepackage{graphicx}
\usepackage{gensymb}
\makeatletter
\usepackage{graphics}\usepackage{subfigure}\usepackage{longtable}\usepackage{pstricks}\usepackage{dcolumn}\usepackage{bm}
\usepackage{amsmath}
\makeatother

\begin{document}

\title{Topotactic transformation of single-crystals: from perovskite to infinite-layer nickelates}


\author{Pascal~Puphal}
\email{puphal@fkf.mpg.de}
\affiliation{Max Planck Institute for Solid State Research, Heisenbergstra{\ss}e 1, 70569 Stuttgart, Germany.}
\author{Yu-Mi Wu}
\affiliation{Max Planck Institute for Solid State Research, Heisenbergstra{\ss}e 1, 70569 Stuttgart, Germany.}
\author{Katrin F{\"u}rsich}
\affiliation{Max Planck Institute for Solid State Research, Heisenbergstra{\ss}e 1, 70569 Stuttgart, Germany.}
\author{Hangoo Lee}
\affiliation{Max Planck Institute for Solid State Research, Heisenbergstra{\ss}e 1, 70569 Stuttgart, Germany.}
\author{Mohammad Pakdaman}
\affiliation{Max Planck Institute for Solid State Research, Heisenbergstra{\ss}e 1, 70569 Stuttgart, Germany.}
\author{Jan A. N. Bruin}
\affiliation{Max Planck Institute for Solid State Research, Heisenbergstra{\ss}e 1, 70569 Stuttgart, Germany.}
\author{J\"urgen Nuss}
\affiliation{Max Planck Institute for Solid State Research, Heisenbergstra{\ss}e 1, 70569 Stuttgart, Germany.}
\author{Y. Eren Suyolcu}
\affiliation{Department of Materials Science and Engineering, Cornell University, Ithaca, New York 14853, USA.}
\author{Peter A. van Aken}
\affiliation{Max Planck Institute for Solid State Research, Heisenbergstra{\ss}e 1, 70569 Stuttgart, Germany.}
\author{Bernhard Keimer}
\affiliation{Max Planck Institute for Solid State Research, Heisenbergstra{\ss}e 1, 70569 Stuttgart, Germany.}
\author{Masahiko Isobe}
\affiliation{Max Planck Institute for Solid State Research, Heisenbergstra{\ss}e 1, 70569 Stuttgart, Germany.}
\author{Matthias~Hepting}
\email{hepting@fkf.mpg.de}
\affiliation{Max Planck Institute for Solid State Research, Heisenbergstra{\ss}e 1, 70569 Stuttgart, Germany.}

\begin{abstract}

Topotactic transformations between related crystal structures are a powerful emerging route for the synthesis of novel quantum materials. Whereas most such ``soft chemistry" experiments have been carried out on polycrystalline powders or thin films, the topotactic modification of single crystals -- the gold standard for physical property measurements on quantum materials -- has proven remarkably difficult. Here, we report the topotactic reduction of La$_{1-x}$Ca$_{x}$NiO$_3$ single-crystals to La$_{1-x}$Ca$_{x}$NiO$_{2+\delta}$ using CaH$_2$ as the reducing agent. The transformation from the three-dimensional perovskite to the quasi-two-dimensional infinite-layer phase was thoroughly characterized by X-ray diffraction, electron microscopy, Raman spectroscopy, magnetometry, and electrical transport measurements. Our work demonstrates that the infinite-layer structure can be realized as a bulk phase in crystals with micrometer-sized single-domains. The electronic properties of these specimens resemble those of epitaxial thin films rather than powders with similar compositions. 

\end{abstract}
\maketitle
\section{Introduction}
\begin{figure*}
 \begin{centering}
\includegraphics[width=2\columnwidth]{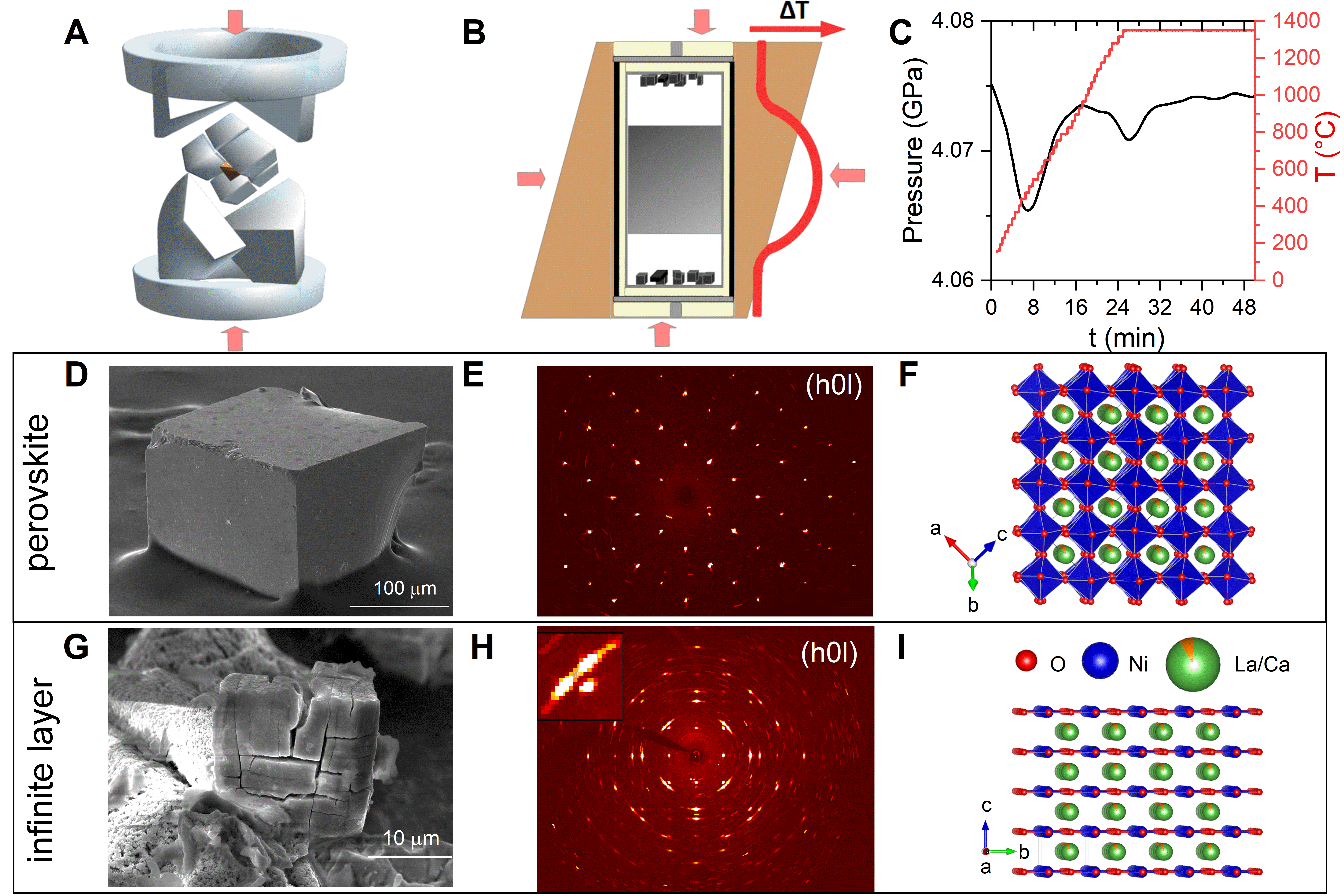}
\par\end{centering}
\caption{\textcolor{black}{\textbf{Synthesis of perovskite nickelate crystals and topotactic reduction.} (\textbf{A}) Schematic drawing of a hydraulic press with a Walker module. The uniaxial force (red arrows) is transferred to an isotropic pressure via a ceramic octahedron (brown) that is embedded in eight tungsten-carbide cubes. (\textbf{B}) Schematic of the cross section of the octahedron with a zirconia crucible inserted (yellow) filled with a Pt capsule (grey line). The salt flux and the nickelate precursor are depicted as white and grey areas in the capsule, respectively. During the growth process, a current is driven through a graphite heater (black line) leading to an external temperature gradient $\Delta T$ (red line). Crystals form at the top and bottom of the capsule (black cubes). (\textbf{C}) Externally applied pressure and temperature of the graphite heater plotted as a function of time. (\textbf{D}) SEM-SE image of an as-grown perovskite single crystal. (\textbf{E}) XRD map of the $(h 0 l)$ planes of an as-grown perovskite single crystal. (\textbf{F}) Crystal structure of an as-grown perovskite crystal with space group $R\bar3c$ according to the refinement of the XRD data in (E). (\textbf{G}) SEM-SE image of a polycrystal after prolonged reduction with CaH$_2$. Separation of domains can be recognized. (\textbf{H}) XRD map of a crystal reduced for a shorter time (see text). The inset shows a set of three reflections corresponding to three orthogonal domains with infinite-layer crystal structure. (\textbf{I}) Crystal structure of an infinite-layer crystal with space group $P4/mmm$ according to the refinement of the XRD data in (H). }}
\label{heating}
\end{figure*}

Synthetic routes to materials in which anions are partially removed, inserted, or exchanged are rapidly gaining attention in solid-state physics and chemistry \cite{Kageyama2018,Jeen2013,Lu2017,Nallagatla2019,Chiu2021}. In particular soft chemistry topotactic reductions \cite{Hayward2002} of perovskite-related transition metal oxides \cite{Anderson1993} have opened up possibilities to prepare entirely new families of functional electronic and magnetic  materials. Prominent examples include the infinite-layer compounds SrFeO$_2$ \cite{Tsujimoto2007} and SrVO$_2$H \cite{DenisRomero2014} obtained via targeted removal of oxygen from the parent perovskites SrFeO$_3$ and SrVO$_3$ using CaH$_2$ as the reducing agent. 

A breakthrough in the field was achieved recently with the discovery of superconductivity in infinite-layer nickelate thin films \cite{Li2019}. More specifically, epitaxial films of Sr or Ca substituted $RE$NiO$_2$ ($RE = $ La, Pr, Nd) obtained via topotactic reduction of the perovskite phase show superconducting transitions below 9 - 15 K \cite{Li2019,Zeng2020,Lee2020,Osada2020,Gu2020,Osada2021,Zeng2021a} and overdamped spin excitations with a bandwidth as large as 200 meV \cite{Lu2021}. These extraordinary properties attracted considerable interest since infinite-layer nickelates are isostructural to cuprate superconductors and were proposed to engender their electronic and magnetic structure \cite{Anisimov1999}. While the relationship between infinite-layer nickelates and cuprates is still an active field of research \cite{Hepting2020,Rossi2020,Goodge2021,Chen2021,Lu2021,Lee2004, Botana2020,Been2021}, it was found that polycrystalline powders with similar compositions show insulating behavior and superconductivity remains elusive \cite{Wang2020, Li2020}. This raises the question whether film-substrate interface effects, including epitaxial strain, polar reconstructions, and interfacial phonons \cite{Reyren2007,Ge2014}, are prerequisites for the emergence of superconductivity in infinite-layer nickelates \cite{Geisler2020, Zhang2020}. 

However, extrinsic factors such as enhanced disorder and diminishing sizes of infinite-layer single-domains in powder grains \cite{Hayward1999,Crespin2005} can possibly lead to physical properties distinct from the genuine bulk properties, which obscures a direct comparison between powders and thin films. 
Thus, single-crystalline specimen with at least micrometer-sized single-domains are highly desired to unveil the intrinsic properties of the infinite-layer phase of nickelates. Moreover, single-crystals allow to take advantage of the application of complementary measurement techniques and can exhibit superior crystalline quality. Yet, synthesis and comprehensive spectroscopic characterization of macroscopic infinite-layer nickelate crystals has not been accomplished, which is also the case for most other topotactically reduced transition metal oxides \cite{Kageyama2018,Parsons2009, Tassel2012, Tsujimoto2007, Ichikawa2012, Seddon2010, DenisRomero2014, Dixon2011, Hayward2002, Chiu2021}.

A major challenge concerning the preparation of infinite-layer nickelate crystals, besides the highly invasive topotactic reduction, is the synthesis of the precursor perovskite phase $RE$NiO$_3$, which requires extreme oxidizing conditions to stabilize the Ni$^{3+}$ valence state. This can be realized through high external oxygen gas pressure \cite{Lacorre1991, Klein2021} or hydrostatic pressure in a Belt-type apparatus, while adding perchlorates as oxidizers \cite{Demazeau1971}. Similarly, salt flux mixtures and perchlorate oxidizers can be employed under hydrostatic pressure and at high temperatures, which yields micrometer-sized $RE$NiO$_3$ single-crystals \cite{Alonso2005,SAITO2003,Alonso2006,Lorenzo2005}. Millimeter-sized LaNiO$_{3}$ \cite{Guo2018, Zhang2017} and PrNiO$_{3}$ \cite{Zheng2019} crystals can be grown by the high oxygen-pressure optical floating zone (OFZ) method \cite{Phelan2019}. However, OFZ grown nickelate crystals might be prone to inclusions of higher-order Ruddlesden-Popper and oxygen-deficient phases \cite{Wang2018,Zheng2019,Zheng2020}. To date, the synthesis of Sr or Ca substituted $RE$NiO$_3$ single-crystals has not been reported.   

Here, we report the high-pressure growth of La$_{1-x}$Ca$_{x}$NiO$_{3}$ single crystals using a 1000 ton press equipped  with  a  Walker  module. We obtained crystals with a typical size of 150 x 150 x 150 \textmu m$^3$ and Ca substitution levels of 10(5)\% in the bulk and 16(3)\% in proximity to the surface. The perovskite crystals were reduced to the infinite-layer phase La$_{1-x}$Ca$_{x}$NiO$_{2+\delta}$ using CaH$_{2}$ and the structural, electronic, and magnetic properties were characterized. Notably, local electron energy-loss spectroscopy reveals close similarities between the electronic structures of our high-quality infinite-layer crystals and thin films. Moreover, we find that the metal-like electrical conductivity of the reduced crystals is reminiscent of weakly hole-doped infinite-layer thin films, which is in stark contrast to previous results on powder samples.

\section{Results and Discussion}
\subsection{High-pressure growth of Ca-substituted perovskite nickelates}

The substitution of the $RE$-ion in $RE$NiO$_{3}$ perovskite nickelates with a divalent ion, such as Sr or Ca, requires Ni to take on an oxidation state even higher than 3+. Thus, for the synthesis of the La$_{1-x}$Ca$_{x}$NiO$_3$ substitution series, we choose a strongly oxidizing environment using salt flux growth with a perchlorate oxidizer under external pressure from a multi-anvil press implemented in a Walker module (Fig.~\ref{heating}A), which is an established route for the synthesis of $RE$NiO$_{3}$ crystals \cite{Alonso2005,SAITO2003,Alonso2006}. To optimize the synthesis process of flux-grown crystals the salt flux (white) and the nickelate precursor (black) were spacially separated, (see Fig.~\ref{heating}B and Materials and Methods for details). The growth is carried out under an external temperature gradient intrinsic to the setup utilizing a graphite heater and dependent on the size of the ampule. Due to the spacial separation of flux and precursor, a transport growth is realized in an external gradient, enabling the growth of relatively large single crystals with sizes limited predominantly by the durability of the crucible. Tests with several different ampule materials and thicknesses showed that Pt is the most suitable crucible material, although it dissolves slowly at the employed temperatures (Fig.~\ref{heating}C). In particular, if the temperature exceeds a certain point, or the holding times are too long for the specific Pt-foil thickness, the flux can dissolve too much of the Pt-ampule and the oxygen pressure is released, which results in the decay of any grown specimen. We found that sufficiently long holding times can be achieved with a  Pt-foil with a thickness of 50 \textmu m. During heating, the underlying chemical processes can be monitored via a change in the external pressure curve (Fig.~\ref{heating}C). First, the flux mixture starts to melt, enabling the decay of the perchlorate. The following release of oxygen then leads to a slow oxidation of the nickelate. Second, at elevated temperatures around 1300\degree C the nickelate is slowly dissolved and the transport growth starts. The dissolved nickelate crystallizes during the holding time at the top and bottom of the Pt-ampule, as depicted in Fig.~\ref{heating}C, after which the ampule is quenched. 

A secondary electron (SE) image of an as-grown crystal acquired with a scanning electron microscope (SEM) is shown in Fig.~\ref{heating}D. Characterization by energy-dispersive X-ray spectroscopy (EDS) (fig.~\ref{SEM}) and single-crystal X-ray diffraction (XRD) indicates that the Ca-substitution level in the obtained crystals is significantly lower than the nominal level of 20\%at, expected from the educts (see Materials and Methods for details). In more detail, our EDS analysis performed on a large number of crystals with sizes up to 200 \textmu m shows that a nominal substitution level of 20 at\% yields La$_{1-x}$Ca$_{x}$NiO$_3$ crystals with $0.06 \leq x \leq 0.16$. Furthermore, we identify concentration gradients towards the center of the crystals (fig.~\ref{EDS}) and slight variations of the substitution level among crystals from the same batch. In particular, we find that the average Ca-substitution on as-grown crystal surfaces is $x = 0.16(3)$, whereas cleaved surfaces, which are representative for the bulk of the crystals, exhibit $x = 0.10(5)$ in average (see also fig.~\ref{EDS}). Moreover, we determine the presence of CaO crystallites on the surface of some crystals with SEM-EDS (fig.~\ref{EDS}) suggesting that the reduced Ca-substitution level is associated with the challenges to stabilize a Ni$^{+3.2}$ oxidation state. We emphasize that the agglomeration of CaO crystallites occurs only on the as-grown crystal surfaces and was not observed on cleaved surfaces, $i.e.$ in the bulk of the crystals. 

From single-crystal XRD (Fig.~\ref{heating}E and fig.~\ref{supp1}), we identify the crystallographic space group of as-grown crystals as $R\bar3c$ (\#167), which is the same space group as reported for single crystalline and powder LaNiO$_3$ \cite{GarciaMunoz1992,Zheng2020,Dey2019}. The refined atomic coordinates and lattice parameters of an as-grown crystal with $x = 0.06(2)$ are given in Table~\ref{table}a. We note that the unit cell dimensions of the Ca-substituted perovskite crystals are slightly larger than those of optical floating zone grown LaNiO$_3$ crystals \cite{Zhang2017,Dey2019}, in spite of the closely similar ionic radii of Ca$^{2+}$ and La$^{3+}$ \cite{Suyolcu2017}. This variation could result from the different synthesis methods and/or different oxygen contents, even though we did not detect any superstructure reflections in XRD (figs.~\ref{supp1} and \ref{supp2}), which are indicative of oxygen deficient phases, such as the La$_{4}$Ni$_{4}$O$_{11}$ and La$_{2}$Ni$_{2}$O$_{5}$ phase \cite{Wang2018,Zheng2020}. 

\begin{figure*}
 \begin{centering}
\includegraphics[width=2\columnwidth]{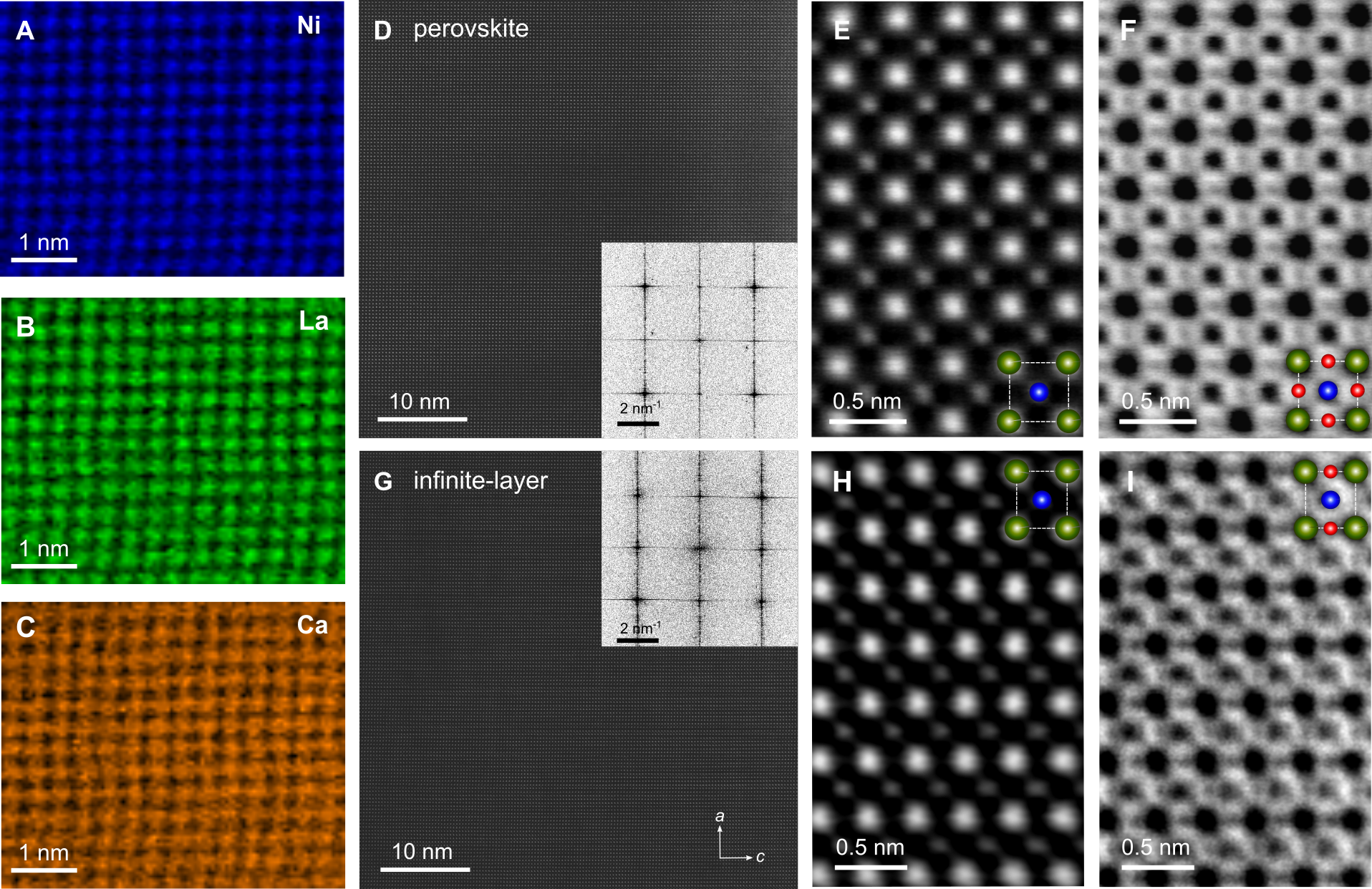} 
\par\end{centering}
\caption{\textcolor{black}{\textbf{Topotactic transformation of the crystal structure.} (\textbf{A} to \textbf{C}) Atomic-resolution STEM-EELS elemental maps of an as-grown perovskite crystal ($x = 0.16$) demonstrating the homogeneous distribution of Ni (A, blue), La (B, green), and Ca (C, orange) atoms. The maps were acquired simultaneously. (\textbf{D} and \textbf{G}) Low-magnification STEM-HAADF images of the perovskite and the reduced crystal, respectively, showing the absence of extended crystallographic defects. Insets correspond to the fast Fourier transformation (FFT) of the HAADF images. (\textbf{E} and \textbf{F}) High-magnification STEM-HAADF and STEM-ABF images of the perovskite crystal. Both images were acquired from the same part of the STEM specimen. The superimposed cartoon indicates a pseudocubic unit cell with different elements highlighted according to their color in the elemental maps (A to C). Oxygen atoms can be identified specifically in the STEM-ABF image (F). (\textbf{H} and \textbf{I}) STEM-HAADF and STEM-ABF images of the reduced crystal, in analogy to (E and F).}}
\label{HAADF}
\end{figure*}

\subsection{Topotactic reduction}

The next step after synthesis of high-quality Ca-substituted perovskite single crystals (Figs.~\ref{heating}A to F) is the topotactic oxygen reduction (Figs.~\ref{heating}G to I). Previously, the reduction process has been investigated in detail for nickelate thin films and polycrystalline powder samples using H$_2$ \cite{Crespin2005} or the reducing agents CaH$_{2}$ or NaH \cite{Hayward1999,Li2019, Zeng2020,Lee2020,Osada2020,Gu2020,Osada2021,Zeng2021a,Ikeda2016,Wang2020,Li2020}. Here, we employed the CaH$_{2}$ variant with spacial separation between reducing agent and sample. Several as-grown crystals were wrapped in aluminum foil and loaded into quartz tubes with approximately 600 mg CaH$_{2}$ powder, which then were evacuated to a high vacuum below 10$^{-7}$ mbar and sealed to ampules with dimensions $\phi_{\text{out}} = 1.7$ mm, $\phi_{\text{in}} = 1.5$ mm, and $h = 10$ cm. We found that after one day of reduction at 300 $^\circ$C the crystals transformed into an intermediate phase (likely the La$_{1-x}$Ca$_{x}$NiO$_{2.5}$ phase). The subsequent reduction to the infinite-layer phase can be accomplished with a significantly longer reduction time of approximately two weeks. The optimal duration of the reduction is individual for each crystal and depends on details, such as crystal size, shape, and most importantly the Ca-substitution level. Overall, we find that an extension of the two weeks time period by a few days did not induce obvious changes in crystals with $0.06 \leq x \leq 0.16$ and sizes between 75 and 200 \textmu m. However, we observed that substantially longer reduction times increase the brittleness of the crystals. After four weeks of reduction, crystals decompose into smaller fragments that mostly exhibit rectangular shapes and are only loosely attached to each other. The SEM-SE image in Fig.~\ref{heating}G displays a representative example for such an \textit{overreduced} polycrystal. Fig.~\ref{heating}H shows a reconstructed map of the $(h0l)$ planes from XRD on a crystal with $x = 0.08(2)$ that was reduced for two weeks. The XRD data can be refined when assuming three orthogonal domains of the tetragonal space group $P4/mmm$ (\#123). Note that the striking difference in the XRD maps of the as-grown perovskite and the reduced crystal (Fig.~\ref{heating}E and H) are due to the rhombohedral and tetragonal symmetries of the respective crystal lattices. The refined $P4/mmm$ symmetry is the same as reported for polycrystalline LaNiO$_{2}$ powder in the infinite-layer phase, which in comparison (Table~\ref{table}b) exhibits slightly smaller in-plane lattice parameters and a larger $c$-axis \cite{Hayward1999}. This difference in lattice parameters can be indicative of a further progressed transformation of the crystals into the infinite-layer phase with less excess oxygen compared to previous powder studies \cite{Hayward1999}. While the presence of three infinite-layer domains is unambiguous from the XRD refinements (Table~\ref{table}b), their average sizes cannot be extracted. However, the shapes of the crystal fragments of the extensively reduced crystal shown in Fig.~\ref{heating}G suggest that they are in the order of tens of \textmu m. Note that substantially smaller domains or micro-twinning would not yield the distinct threefold peak splitting observed in the XRD maps (see inset in Fig.~\ref{heating}H).         

\begin{figure*}
 \begin{centering}
\includegraphics[width=2.0\columnwidth]{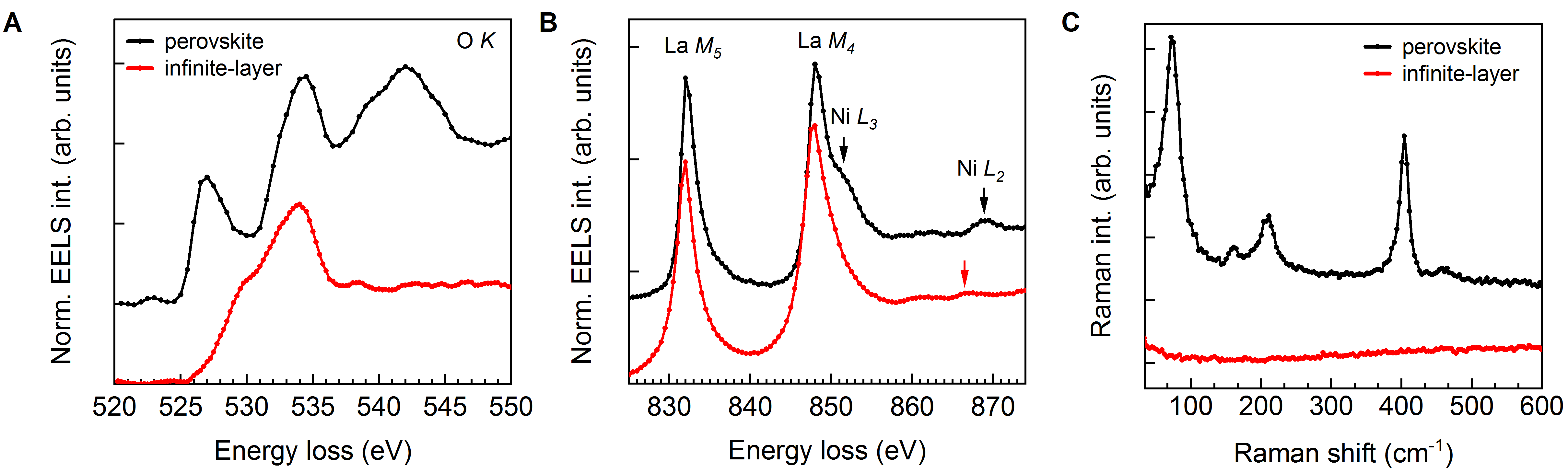} 
\par\end{centering}
\caption{\textcolor{black}{\textbf{Spectroscopic characterization.} (\textbf{A}) STEM-EELS of the O-$K$ near-edge fine structure of an as-grown perovskite (black line) and a reduced (red) crystal, respectively. The O $K$-edge pre-peak of the perovskite phase at $\sim$527 eV is associated with hybridized Ni-O states and vanishes upon reduction, whereas the hole-doped infinite-layer phase exhibits enhanced spectral weight around $\sim$529 eV. (\textbf{B}) STEM-EELS across the La $M_{5,4}$ ($\sim$832 eV and $\sim$847 eV) and Ni $L_{3,2}$ ($\sim$852 eV and $\sim$868 eV) edges. Black and red arrows indicate the Ni $L_{3,2}$ peak positions of the perovskite and the infinite-layer phase, respectively. Spectra are normalized to the intensity of the La $M_5$ peak. (\textbf{C}) Raman spectra of an as-grown perovskite (black line) and a reduced (red) crystal, respectively. Measurements were performed at $T = 300$ K with a laser wavelength $\lambda = 632.8$ nm. Spectra in all panels are vertically offset for clarity.}}
\label{Kedge} 
\end{figure*}

\begin{table}
\caption{\label{refinement} (a) Refined lattice parameters and atomic coordinates of an as-grown crystal. The refinement was performed in the rhombohedral space group $R\bar3 c$ (hexagonal axes). A Ca-subtitution level of $x = 0.06(2)$ was extracted. The reliability factor is $\chi^{2} = 1.496$. (b) Refined parameters of a reduced crystal in the tetragonal space group $P4/mmm$. A Ca-subtitution level of $x = 0.08(2)$ was extracted. The reliability factor is $\chi^{2} = 1.227$.
$U$ gives the isotropic expansion and Occ. the atom occupation.}

\begin{tabular}{cccccc}
\hline 
\multicolumn{6}{c}{{(a) $\vert$ $a,b=5.464(3)$ $\text{\AA}$, $c=13.166(9)$
$\text{\AA}$}}\\
\hline 
{Atom } & {$x/a$} & {$y/b$} & {$z/c$} & {$U [\text{\AA}^2]$} & {Occ.}\\
\hline 
{La (6a) } & {0 } & {0 } & {0.25 } & {0.0097(4) } & {0.937(15)}\\
{Ca (6a) } & {0 } & {0 } & {0.25 } & {0.0097(4) } & {0.063(15)}\\
{Ni (6b) } & {0 } & {0 } & {0 } & {0.0072(6) } & {1}\\
{O (18e) } & {0.4559(9) } & {0 } & {0.25 } & {0.0163(8) } & {1}\\
 &  &  &  &  & \\
\multicolumn{6}{c}{{(b) $\vert$ $a,b=3.9637(9)$ $\text{\AA}$, $c=3.3663(10)$
$\text{\AA}$}}\\
\hline 
{Atom } & {$x/a$} & {$y/b$} & {$z/c$} & {$U [\text{\AA}^2]$} & {Occ.}\\
\hline 
{La (1d) } & {0.5 } & {0.5 } & {0.5 } & {0.0184(5) } & {0.921(18)}\\
{Ca (1d) } & {0.5 } & {0.5 } & {0.5 } & {0.644(10) } & {0.079(18)}\\
{Ni (1a) } & {0 } & {0 } & {0 } & {0.0190(8) } & {1}\\
{O (2f) } & {0 } & {0.5 } & {0 } & {0.0204(19) } & {1}\\
\end{tabular}
\label{table}
\end{table}

\subsection{Electron microscopy investigation of the topotactic transformation}

The local crystal structure and the elemental distribution in the perovskite and infinite-layer crystals was investigated by scanning transmission electron microscopy (STEM) including electron energy-loss spectroscopy (EELS). Figures~\ref{HAADF} A to C show the simultaneously acquired atomic-resolution STEM-EELS elemental maps of Ni, La, and Ca in a perovskite crystal 
recorded using the Ni $L_{3,2}$, La $M_{5,4}$ and Ca-$L_{3,2}$ edges, respectively. The maps signify  a homogeneous distribution of the elements, including the substitution species Ca. The Ca substitution amount in the STEM specimen was determined with STEM-EDS as $x \sim 0.16$ and remains unchanged within a region of least 2 $\mu$m in proximity to the top surface of the STEM specimen (fig.~\ref{domain}). The substitution level is consistent with an EDS measurement in a SEM on the surface of the as-grown crystal. However, XRD on a fragment of the same crystal revealed $x = 0.08(2)$, thus suggesting a strong substitution gradient towards the bulk of the crystal on length scales of several tens of micrometer. The low-magnification STEM high-angle annular dark-field (HAADF) images in Figs.~\ref{HAADF}D and G reveal the absence of crystallographic defects on length scales of several tens of nanometer for both, the perovskite and infinite-layer phase. The observed high crystalline quality is in line with the single crystal XRD characterization (Figs.~\ref{heating}E and H). Note that the two STEM specimens were prepared from the same crystal, before and after the reduction. Importantly, the similarity of the crystal lattices in Figs.~\ref{HAADF}D and G indicates that the integrity of the $A$-site cation sublattice remains intact during the reduction process. Profound changes of the $A$-site cation sublattice can be extracted from the fast Fourier transformed (FFT) amplitudes of the images, which are displayed as insets in Figs.~\ref{HAADF}D and G. Along the horizontal direction of the FFT images the distance between features increases upon reduction, corresponding to a contraction of the $c$-axis lattice parameter in real space, which is consistent with the removal of oxygen ions. In the vertical direction, the distances remain almost unchanged. Specifically, the ratio between lattice parameters $c/a \sim 0.85$ calculated from the FFT maxima for the reduced crystal is similar to the result obtained with XRD (Table~\ref{table}b). The trend of a collapse of the $c$-axis in real space can also be observed in the high magnification STEM-HAADF images in Figs.~\ref{HAADF}E and H. 

Detailed information about the distribution of oxygen ions in the lattice can be provided by STEM Annular Bright Field (ABF) imaging (Figs.~\ref{HAADF}F and I). While Ni in the cross-sectional STEM-ABF image of the perovskite crystal (Fig.~\ref{HAADF}F) is coordinated with four oxygen ions, we observe in the reduced crystal that Ni is  coordinated with two oxygen ions for the most part of the specimen (Fig.~\ref{HAADF}I), which signals the presence of the infinite-layer structure. The resulting orientation of the NiO$_2$ planes within the infinite-layer structure is indicated in Fig.~\ref{HAADF}I. However, we note that in some regions in Fig.~\ref{HAADF}I also the supposedly non-occupied oxygen positions exhibit a slightly dark contrast, which suggests that some apical oxygens withstand the reduction. Furthermore, a subtle variation of the contrast among the occupied oxygen positions of the NiO$_2$ planes can be indicative of an occational depletion within the NiO$_2$ planes upon reduction. Overall, we stress that the STEM-HAADF and ABF images of the reduced crystal in Figs.~\ref{HAADF}G to I confirm a clear, high quality infinite-layer structure, while local non-stoichiometries of the oxygen ions can be identified in Fig.~\ref{HAADF}I. In consequence, we conclude that the infinite-layer structure is realized in our crystals within volumes of several cubic micrometers, whereas for thin films the thickness of the infinite-layer phase was reported to be less than 10 nm \cite{Lee2020}. Moreover, Ruddlesden-Popper phase inclusions or stacking faults in the crystal lattice, which have typically been identified in thin films \cite{Lee2020},  were not observed in our crystals.

\subsection{Electronic structure of reduced crystals}

Having established the details of the crystal structure, we proceed with an investigation of the electronic structure of our samples. The characteristic multiband electronic structure of infinite-layer nickelate thin films has been revealed in previous X-ray absorption spectroscopy (XAS) and STEM-EELS studies  \cite{Hepting2020,Goodge2021,Rossi2020,Zeng2021,Ortiz2021}. Specifically the degree of Ni-O hybridization can be deduced from the near-edge fine structure of the O $K$-edge. In the as-grown perovskite crystal ($x_{\text{EDS}} \sim 0.16$), we detect a pronounced pre-peak at $\sim$527 eV  (Fig.~\ref{Kedge}A), characteristic of a strong hybridization between ligand O 2$p$ and Ni 3$d$ states in $RE$NiO$_{3}$ nickelates \cite{Abbate2002}. Noticeably, the spectral weight at this energy vanishes after reduction, which is in line with XAS and STEM-EELS studies on reduced films, where the lack of the pre-peak was interpreted as the absence of Ni-O hybridization, while a Mott-Hubbard character emerges \cite{Hepting2020,Goodge2021,Rossi2020}. Concomitant with the vanishing of the 527 eV pre-peak, a recent STEM-EELS study reported the appearance of additional spectral weight around 528 eV as a function of hole-doping, which could be reminiscent of the Zhang-Rice singlet peak in  isostructural cuprates \cite{Goodge2021}. In fact, we observe a similar feature centered around 529 eV in our reduced and Ca-substituted crystal. In addition, we notice that a broad feature of the perovskite spectrum centered around 542 eV, which is likely associated with hybridized Ni 4$sp$ states \cite{Abbate2002}, vanishes upon reduction. A similar trend was observed in NdNiO$_{3}$/NdNiO$_{2}$ STEM-EELS spectra \cite{Goodge2021}, which calls for future STEM-EELS and XAS studies to clarify the origin of this behavior.   

Further insights into the electronic structure can be gained from STEM-EELS across the La $M_{5,4}$ and Ni $L_{3,2}$ edges. Fig.~\ref{Kedge}B shows that upon reduction the La $M_{5}$ peak remains unchanged within the experimental error, as expected for an empty 4$f$ shell of La$^{3+}$ and an unchanged valence state \cite{Ortiz2021}. 
The La $M_{4}$ and Ni $L_{3}$-edges overlap strongly, but nevertheless a redistribution of spectral weight of the Ni $L_{3}$-edge towards lower energy can be recognized for the infinite-layer phase, which is consistent with XAS studies \cite{Hepting2020,Zeng2021,Ortiz2021}. The Ni $L_{2}$-edge is well-separated from other features and shifts towards lower energies, as expected from the lowering of the 3+ valence state of Ni in the reduced phase.

\subsection{Lattice dynamics}

While STEM-EELS provided information about the local electronic structure, next we use Raman spectroscopy to probe lattice dynamics averaged over length scales of several micrometer. Fig.~\ref{Kedge}C shows the Raman spectra of a perovskite and an infinite-layer crystal ($x = 0.06(3)$). As expected \cite{Gou2011}, five Raman active modes (one A$_{1g}$ and four E$_{g}$) are observed for the perovskite crystal with space group $R\bar3c$ (Table~\ref{table}a). The Raman frequencies are similar to those reported for LaNiO$_3$ films \cite{Chaban2010, Hepting2015}. For the reduced sample in the infinite-layer structure and space group $P4/mmm$ (Table~\ref{table}b) no first-order Raman active modes are expected \cite{Burns1989, Castro2013}. Accordingly, we do not detect any phonon-like feature in the Raman spectra of crystals that are reduced to such an extent that their average crystal structure can be assigned to the infinite-layer phase (Fig.~\ref{Kedge}C).

\subsection{Magnetic susceptibility and electronic transport}

\begin{figure*}
 \begin{centering}
\includegraphics[width=1.5\columnwidth]{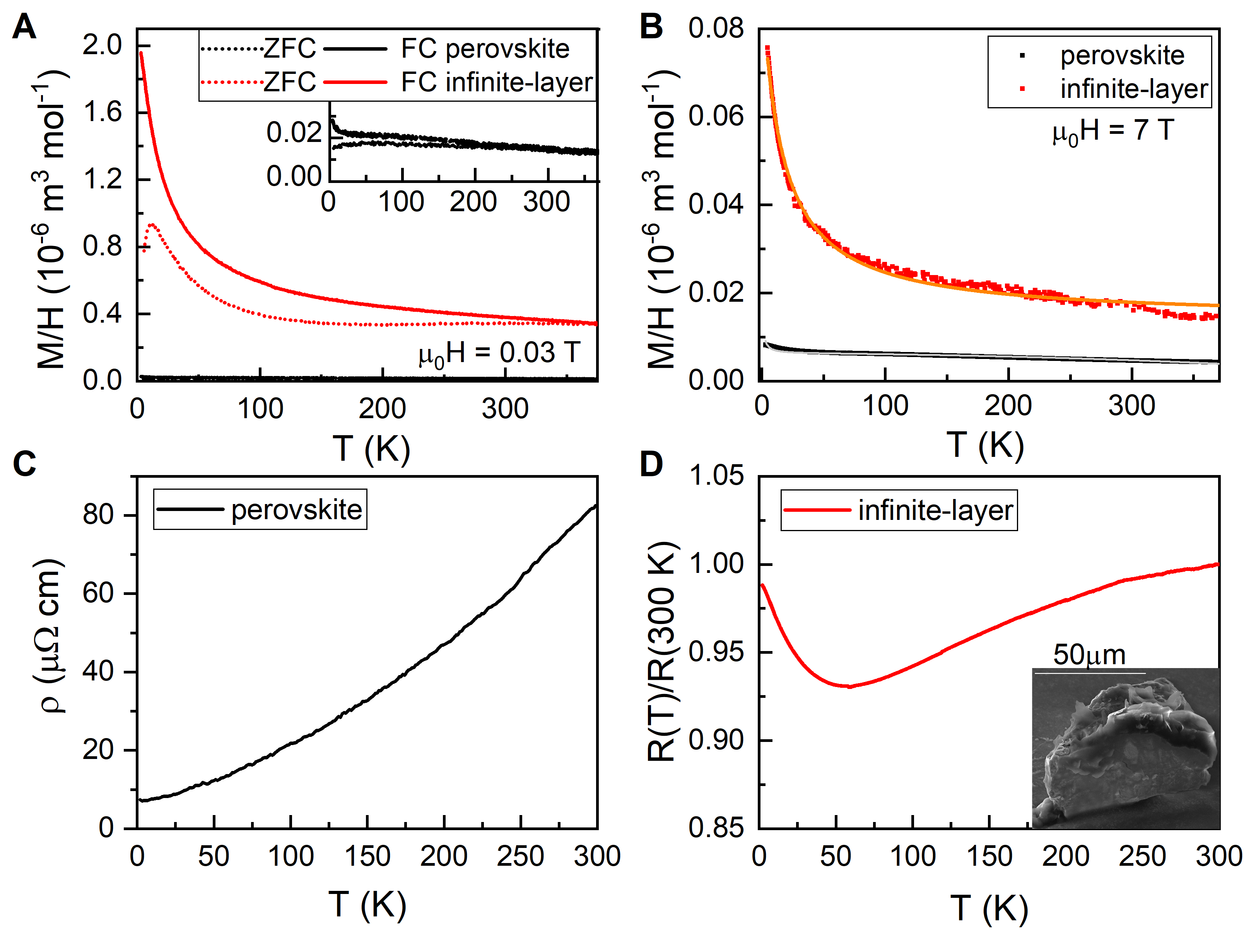} 
\par\end{centering}
\caption{\textcolor{black}{\textbf{Magnetic properties and electrical transport.} (\textbf{A}) Magnetic susceptibility of an as-grown perovskite (black) and a reduced (red) crystal measured upon zero field cooling (ZFC, dashed lines) and field cooling (FC, solid lines) in a small external field of 0.03 T. (\textbf{B}) Susceptibility in a strong field of 7 T. The solid gray and orange lines are fits with a Curie-type law (see text). (\textbf{C}) Resistivity of a perovskite single crystal ($x = 0.07(2) $). (\textbf{D}) Resistance of a reduced crystal ($x = 0.08(2)$) normalized to the room-temperature value. The inset shows an SEM-SE image of a fragment of the infinite-layer crystal utilized in both, the electrical transport and STEM measurements.}}
\label{susc}
\end{figure*}

Figures~\ref{susc}A and B show the magnetic susceptibility of an as-grown perovskite and a reduced crystal measured in small and large external magnetic fields, respectively. As displayed in Fig.~\ref{susc}B, we observe paramagnetic behavior for our Ca-doped perovskite single crystal ($x = 0.07(2)$), similar to polycrystalline LaNiO$_{3}$ \cite{Zhou2014} and OFZ-grown LaNiO$_{3}$ single crystals \cite{Wang2018,Zheng2020}. Signatures of a magnetic transition around 170 K as reported for some OFZ crystals \cite{Zhang2017} and LaNiO$_{3}$ with La$_2$Ni$_2$O$_{5}$ and/or Ruddlesden-Popper inclusions \cite{Zheng2020} were not detected. This is consistent with the absence of such lattice defects in the STEM analysis (Fig.~\ref{HAADF}).

Remarkably, measurements in small fields (Fig.~\ref{susc}A) reveal a bifurcation between the field-cooled (FC) and zero-field-cooled (ZFC) susceptibility for the reduced crystal that persists up to the highest measured temperature (375 K), but is not present in measurements in strong fields (Fig.~\ref{susc}B). In addition, the ZFC curve in small fields shows a cusp-like feature around 10 K. Such behavior is typical for spin glasses \cite{Huangfu2020} and is likely not an intrinsic property of the infinite-layer phase, but can originate from local oxygen non-stoichiometries, which were identified with STEM-ABF (Fig.~\ref{HAADF}I).  Furthermore, a small bifurcation between the FC and ZFC susceptibility also exists in case of the non-reduced perovskite crystal (inset in Fig.~\ref{susc}A). Moreover, a bifurcation and glassy properties were recently detected in perovskite Sm$_{1-x}$Sr$_x$NiO$_3$ powders \cite{He2021}. This supports the notion that the spin glass behavior is not an intrinsic propoerty of the infinite-layer phase. Instead, a small number of oxygen vacancies due to challenges associated with the stabilization of a Ni valence state higher than 3+ in the Ca- or Sr-substituted perovskite phase and local oxygen non-stoichiometry in the infinite-layer phase presumably lead to the observed bifurcations. Alternatively, a FC-ZFC splitting can be due to ferromagnetic impurities, such as elemental Ni particles, which were reported for reduced nickelate powder samples \cite{Hayward1999}. However, the XRD, STEM, and EDS characterization of our crystals did not indicate the presence of any such impurities.

Along the lines of paramagnetic LaNiO$_{3}$, we fit the susceptibility of the perovskite crystal by a Curie-type law $\chi(T) = \chi_{0}(1-aT^2) + C/T$ \cite{Zhou2014}, which includes Pauli and van Vleck paramagnetism as well as Landau and core diamagnetism. $\chi_{0}$ is a temperature-independent constant and $C$ the Curie constant. The fit yields $a=2.66(2)\cdot10^{-6}$ 1/K$^2$, $C=7.1(1)\cdot10^{-4}$ emu K/mol and $\chi_{0}=4.922(7)\cdot10^{-4}$ emu/mol, which is similar to LaNiO$_{3}$ \cite{Zhou2014}. Hence, a Ca substitution of $x = 0.08(2) $ does not substantially affect the magnetic correlations in the perovskite case. The enhanced signal of the infinite-layer polycrystal is fitted with a Curie-Weiss law $\chi(T) = \chi_{0} + C/(T-\theta _W)$ and we obtain $C=0.099(2)$  emu K/mol, $\chi_{0}=0.0011(1)$ emu/mol, and a Curie-Weiss-temperature $\theta _W=-16.3(5)$ K.

The electrical transport properties of a perovskite and a reduced crystal are displayed in Figs.~\ref{susc}C and D, respectively. For the perovskite crystal ($x = 0.07(3)$), the expected metallic behavior is observed (Fig.~\ref{susc}C) and we determine a residual-resistivity ratio (RRR) of 11. This value is slightly larger than the reported RRR of $\sim 9.6$ of OFZ-grown LaNiO$_{3}$ single crystals \cite{Zhang2017} and can be related to the Ca substitution and/or the increased quality of our flux-grown crystals. While the resistivity of the perovskite crystal (Fig.~\ref{susc}C) was determined via a four-point measurement, the reduced crystal was measured in two-point configuration (Fig.~\ref{susc}D) owing to the smaller size of reduced crystals (see inset in Fig.~\ref{susc}D), which tend to break into smaller pieces along the domain boundaries (Fig.~\ref{heating}G). The resistance $R(T)$ of a reduced crystal ($x=0.08$) normalized to the room temperature resistance is shown in Fig.~\ref{susc}D. Strikingly, we find that the resistance decreases with decreasing temperature and exhibits a subtle upturn below $\sim 50$ K. This is in stark contrast to powder samples, which show insulating behavior at all temperatures \cite{Wang2020, Li2020}. Instead, the electrical resistance of our reduced crystal is reminiscent of lightly hole-doped non-superconducting $RE$NiO$_2$ thin films \cite{Li2019,Zeng2020,Lee2020,Osada2020,Gu2020,Osada2021,Zeng2021a} and underdoped cuprates \cite{Keimer1992} that exhibit metallic behavior at high temperatures and a weakly insulating upturn at low temperatures. The upturn in nickelate films was recently interpreted as a signature of strong electron correlations \cite{Hsu2021}. We note that the upturn in Fig.~\ref{susc}D coincides approximately with the onset temperature of the strong upturn in the magnetic susceptibility (Fig.~\ref{susc}B). 
Future studies will be required to clarify the origin of the discrepancy between the electrical transport properties of powders and crystals, which could be due to inferior crystalline quality, impurities \cite{Hayward1999,Wang2020,Li2020,He2021}, domain/grain boundaries, or enhanced hydrogen intercalation \cite{Si2020} in the powders.


In summary, we synthesized perovskite La$_{1-x}$Ca$_{x}$NiO$_3$ crystals in the perovskite phase via a perchlorate-chlorate flux mixture in an external temperature gradient growth under extreme pressures. The perovskite crystals were successfully reduced to the infinite-layer phase La$_{1-x}$Ca$_{x}$NiO$_{2+\delta}$ with three orthogonally oriented crystallographic domains. Excessive reduction increased detachment between domains. Nevertheless,  micrometer-sized domains remained robust and showed excellent crystalline quality with homogeneous Ca distribution and no detectable defects on length scales exceeding the thickness of previously reported films by orders of magnitude. The oxygen sublattices in the reduced crystals are consistent with the infinite-layer structure, but exhibited local non-stoichiometries of residual apical oxygen and/or vacancies within the NiO$_2$ planes. To date, detailed information about the corresponding local oxygen stoichiometry in thin films is sparse. Yet, the metal-like electrical transport observed for the crystals suggested a close similarity to weakly hole-doped thin films. In consequence, with somewhat higher Ca substitution in the bulk, infinite-layer crystals are a promising candidate for hosting superconductivity. In a broader context, our work signifies that topotactic reductions can be applied to bulk single-crystalline specimen. To the best of our knowledge, this was previously only realized for perovskite-derived layered materials \cite{Zhang2017N}. Hence, our successful reduction of the perovskite to the infinite-layer structure provides new perspectives for the transformation of three-dimensional to quasi-two-dimensional materials. In particular, we anticipate that topotactic reductions are feasible for a wide variety of perovskite oxides with distinct morphologies, including nanowires, fibers, and cubes \cite{Zhao2012,Zhang2014}. While such perovskites are already widely employed as functional materials for energy storage and conversion \cite{Kubicek2017}, catalysis \cite{Chen2015,Hong2015}, and ionic conduction \cite{Coors2004,Sun2018}, they might show superior performances or new functionalities after topotactic transformation.

\section{Materials and Methods}
\subsection{Sample preparation}

For the high pressure growth La$_{2}$O$_{3}$ powder (0.39056 g, 3 mmol, Alfa  Aesar 99.99 \%) was dried at 1000$^{\circ}$C for one day. Subsequently, we  mixed and ground the powder in a 0.8 : 0.2 : 1 molar ratio with CaO (0.03361 g, 3 mmol, Sigma Aldrich 99.9 \%) and NiO (0.22384 g, 3 mmol, Alfa  Aesar 99.998 \% metal powder) to obtain stoichiometric  La$_{0.8}$Ca$_{0.2}$NiO$_{3}$ single crystals. The spacially separated salt flux was prepared by mixing a molar ratio of 0.1 : 0.3 : 0.6 of KCl (0.02234 g, 3 mmol, Roth 99.5\%), KClO$_{4}$ (0.12456 g, 3 mmol, Sigma-Aldrich 99.999\% metals basis) and NaCl (0.10509 g, 3 mmol, Sigma-Aldrich 99.999\% metals basis). The mixtures were sealed in a platinum-foil ampule ($\phi=7$ mm, $h=10$ mm)  in a sandwich structure of flux - nickelate mixture - flux.  The ampule was heated to $1380^{\circ}$C for 1-2h under a pressure of 4 GPa in a Max Voggenreiter mavo press LP 1000-540/50 equipped with a Walker module for 32 mm WC-cubes and subsequently quenched to room temperature.

\subsection{Single-crystal X-ray diffraction}
Since only very tiny crystal pieces turned out to be suitable for single-crystal X-ray diffraction, perovskite crystals were broken under high viscosity oil. A small piece was mounted with some grease on a loop made of Kapton foil (Micromounts$^{TM}$, MiTeGen, Ithaca, NY). Diffraction data were collected at room temperature with a SMART APEXI CCD X-ray diffractometer (Bruker AXS, Karlsruhe, Germany), using graphite-monochromated Mo-K$_{\alpha}$ radiation ($\lambda=0.71073\,$\AA). The reflection intensities were integrated with the SAINT subprogram in the Bruker Suite software package \cite{Brukersuite}. For the rhombohedral perovskite La$_{1-x}$Ca$_{x}$NiO$_3$, a multi-scan absorption correction was applied using SADABS. Crystals of the tetragonal, reduced infinite-layer nickelate, showed systematic twinning by reticular merohedry. The threefold axes of the (pseudo)cubic perovskite structure become twinning elements, and the three twin domains are related by transformation matrices: (1 0 0) (0 1 0) (0 0 1); (0 1 0) (0 0 1) (1 0 0); (0 0 1) (1 0 0) (0 1 0). To handle this, the reflection intensities were integrated with the help of the orientation matrices of all three twin-domains, and a multi-scan absorption correction was applied using TWINABS \cite{Krause2015}. Both structures were solved by direct methods and refined by full-matrix least-square fitting with the SHELXTL software package 
\cite{Sheldrick2015}.
Further details of the crystal structure analysis may be obtained from the Fachinformationszentrum Karlsruhe, 76344 Eggenstein-Leopoldshafen, Germany (Fax: +49-7247-808-666; E-Mail: crysdata@fiz-karlsruhe.de, http://www.fiz-karlsruhe.de/request for de-posited data.html) on quoting the depository numbers CSD-2086960 and CSD-2086915, respectively.

\subsection{Scanning transmission electron microscopy}
Electron-transparent TEM specimens of the as-grown and reduced sample were prepared on a Thermo Fischer Scios I focussed ion beam (FIB) using the standard liftout method. The lateral dimensions of the specimens were 20 \textmu m $\times$ 1.5 \textmu m with thicknesses between 50 - 100 nm. Energy-dispersive X-ray spectra (EDS) were recorded with an NORAN System 7 (NSS212E) detector in a Tescan Vega  (TS-5130MM) scanning electron microscope (SEM).
High-angle annular dark-field imaging (HAADF), annular bright-field (ABF) and electron energy-loss spectroscopy (EELS) were recorded by a probe-aberration-corrected JEOL JEM-ARM200F STEM equipped with a cold field-emission electron source, a probe Cs-corrector (DCOR, CEOS GmbH), and a Gatan K2 direct electron detector with a large solid-angle JEOL Centurio SDD-type EDS detector was used at 200 kV. STEM imaging and EDS and EELS analyses were performed at probe semi-convergence angles of 20 mrad and 28 mrad, resulting in probe sizes of 0.8 \text{\AA} and  1.0 \text{\AA}, respectively. Collection angles for STEM-HAADF and ABF images were 75-310 mrad and 11-23 mrad, respectively. To improve the signal-to-noise ratio of the STEM-HAADF and ABF data while minimizing sample damage, a high-speed time series was recorded (2 \textmu s per pixel), and was then aligned and summed. A collection semi-angle of 111 mrad was used for EELS investigations. A 0.5 eV/ch dispersion with an effective energy resolution of $\sim$1 eV was used for overall chemical profiling and 0.1 eV/ch dispersion with an effective energy resolution of $\sim$0.5 eV was chosen particularly for the O $K$ edges.

\subsection{Raman spectroscopy}
The Raman measurements were performed with a Jobin-Yvon LabRam HR800 single-grating (1800) Raman spectrometer using the 632.8 nm excitation line of a HeNe laser. Spectra were taken at 300 K in backscattering geometry with parallel and crossed polarization, respectively. The data displayed in Fig.~\ref{Kedge}C correspond to the sum of the spectra recorded with the two polarization configurations. 

\subsection{Physical properties measurements}
Magnetic susceptibility measurements were carried out in a range of
1.8 - 400\,K and 0 - 7\,T using a Quantum Design Magnetic Property
Measurements System (MPMS). We used silver paint to contact the samples for resistivity and resistance measurements, which were carried out using the standard resistivity option of a Physical Property Measurements System (PPMS). As explained in the main text, the perovskite crystal was measured in a four-probe geometry, and a two-contact resistance measurement was employed for the substantially smaller infinite-layer crystal. We note that our two-contact measurement of the infinite-layer crystal cannot determine intrinsic transport properties quantitatively, as this measurement configuration adds contributions from electrical contact resistances. Nevertheless, assuming that the temperature dependence of the contact resistances is small and/or monotonic, the $R(T)$/$R(300 K)$ data in Fig.~\ref{susc}D are qualitatively representative for the temperature-dependence of the intrinsic electrical transport.

\section{Supplementary materials}
Supplementary material for this article is available at [...]

\section{References and notes}

\bibliographystyle{apsrev4-1}
\addcontentsline{toc}{section}{\refname}\nocite{*}
\bibliography{nickelates}

\bigskip\noindent\textbf{Acknowledgments:} We thank Graham McNally for fruitful discussions about high pressure synthesis. The use of facilities and resources of the Quantum Materials Department of Prof. Hidenori Takagi at MPI-FKF is gratefully acknowledged. \textbf{Funding:} This project has received funding from the European Union's Horizon 2020 research and innovation programme under Grant Agreement No. 823717-ESTEEM3. \textbf{Author contributions:} M.H., P.P., M.I., and B.K. conceived the project. P.P. and M.I. grew the perovskite crystals and P.P. carried out the reduction. The Raman measurements were executed by H.L. under the supervision of K.F. and M.H. The resistivity measurements were conducted by M.P. and J.B and discussed with K.F. The single crystal diffraction and data analysis was performed by J. N. The STEM measurements were executed by Y.-M.W., and Y.E.S. and P.A. van A. supervised the analysis. P.P. and M.H. characterized the samples with SQUID. P.P. and M.H. wrote the manuscript with input from all authors. \textbf{Competing interests:} The authors declare that they have no competing interests. \textbf{Data and materials availability:} All data needed to evaluate the conclusions in the paper are present in the paper and/or the Supplementary Materials. Additional data related to this paper may be requested from the authors.


\newpage

\section{Supplementary Material}

\renewcommand{\thefigure}{S\arabic{figure}}

\setcounter{figure}{0}

\subsection{Details of the single-crystal X-ray diffraction}

\begin{table*}
\caption{Crystal data, data collection and refinement details at 298 K. $^{[a]}$
Further details of the crystal structure investigations may be obtained
from the Fachinformationszentrum Karlsruhe, D-76344 Eggenstein-Leopoldshafen,
Germany, on quoting the depository number (http://www.fiz-karlsruhe.de).}

\begin{tabular}{|c|c|c|}
\hline 
 & Ca$_{0.06}$La$_{0.94}$NiO$_{3}$ & Ca$_{0.08}$La$_{0.92}$NiO$_{2}$\tabularnewline
\hline 
\hline 
Formula weight  & 239.36  & 221.71 \tabularnewline
\hline 
Crystal system  & trigonal  & tetragonal \tabularnewline
\hline 
Space group (no.), Z & $R\bar3c$ (167), 6 & $P4/mmm$ (123), 1\tabularnewline
\hline 
Lattice parameters / $\text{\AA}$ & a = 5.464(3)  & a = 3.9637(9) \tabularnewline
\cline{2-3} 
 & c = 13.166(9)  & c = 3.3663(10) \tabularnewline
\hline 
V /\AA$^{3}$ & 340.4(4)  & 52.89 \tabularnewline
\hline 
$\rho_{xray}$ /g$\cdot$ cm$^{-3}$  & 7.005  & 6.961 \tabularnewline
\hline 
Crystal size /mm$^{3}$  & 0.03\texttimes 0.03\texttimes 0.01  & 0.05\texttimes 0.03\texttimes 0.02 \tabularnewline
\hline 
Diffractometer  & \multicolumn{2}{c|}{SMART APEX-I, Bruker AXS }\tabularnewline
\hline 
X-ray radiation, $\lambda$ / \AA  & \multicolumn{2}{c|}{Mo-K$_\alpha$, 0.71073 }\tabularnewline
\hline 
Absorption correction  & Multi-scan, SADABS  & Multi-scan, TWINABS \tabularnewline
\hline 
2$\theta$ range /\textdegree{}  & 10.614 \ensuremath{\le} 2$\theta$ \ensuremath{\le} 70.230  & 10.288 \ensuremath{\le} 2$\theta$ \ensuremath{\le} 70.536 \tabularnewline
\hline 
Index ranges  & \textendash 8 \ensuremath{\le} h \ensuremath{\le} 8, \textendash 8
\ensuremath{\le} k \ensuremath{\le} 8, \textendash 20 \ensuremath{\le}
l \ensuremath{\le} 20  & 0 \ensuremath{\le} h \ensuremath{\le} 4, 0 \ensuremath{\le} k \ensuremath{\le}
6, 0 \ensuremath{\le} l \ensuremath{\le} 5 \tabularnewline
\hline 
Reflections collected  & 1542  & 2342 \tabularnewline
\hline 
Twin volume fractions, V$_{1}$-V$_{3}$  & \textemdash{}  & 0.40(3), 0.31(3), 0.29(5) \tabularnewline
\hline 
Data, $R_{int}$  & 174, 0.034 & 95, 0.116\tabularnewline
\hline 
No. of parameters  & 12 & 11\tabularnewline
\hline 
Transmission: $t_{min}$, $t_{max}$ & 0.180, 0.272 & 0.102, 0.272\tabularnewline
\hline 
$R_{1}$ {[}F$^{2}>2\sigma$(F$^{2}$){]}  & 0.023 & 0.047\tabularnewline
\hline 
w$R$(F$^{2}$)  & 0.084 & 0.112\tabularnewline
\hline 
Deposition no.$^{[a]}$ & 2086960 & 2086915\tabularnewline
\hline 
\end{tabular}
\end{table*}

\begin{figure*}[h]
 \begin{centering}
\includegraphics[width=2\columnwidth]{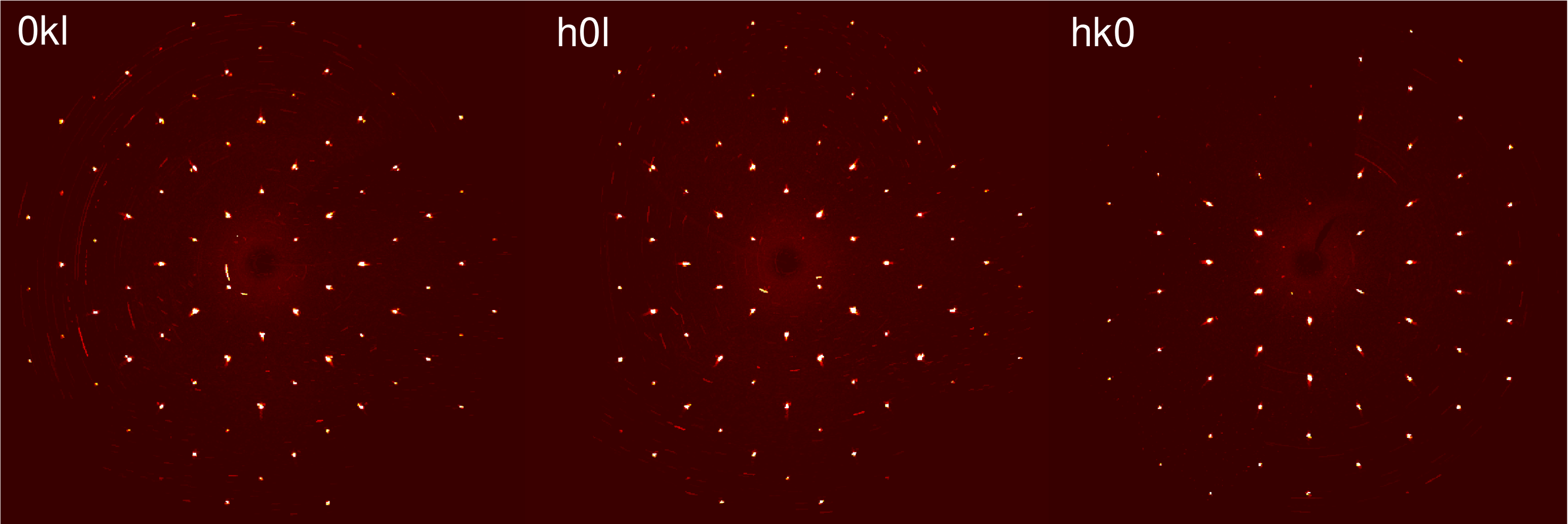}
\par\end{centering}
\caption{{\label{supp1}XRD maps of the (0kl), (h0l) and (hk0) planes of the investigated perovskite crystal.}}
\end{figure*}

\begin{figure*}[h]
 \begin{centering}
\includegraphics[width=2\columnwidth]{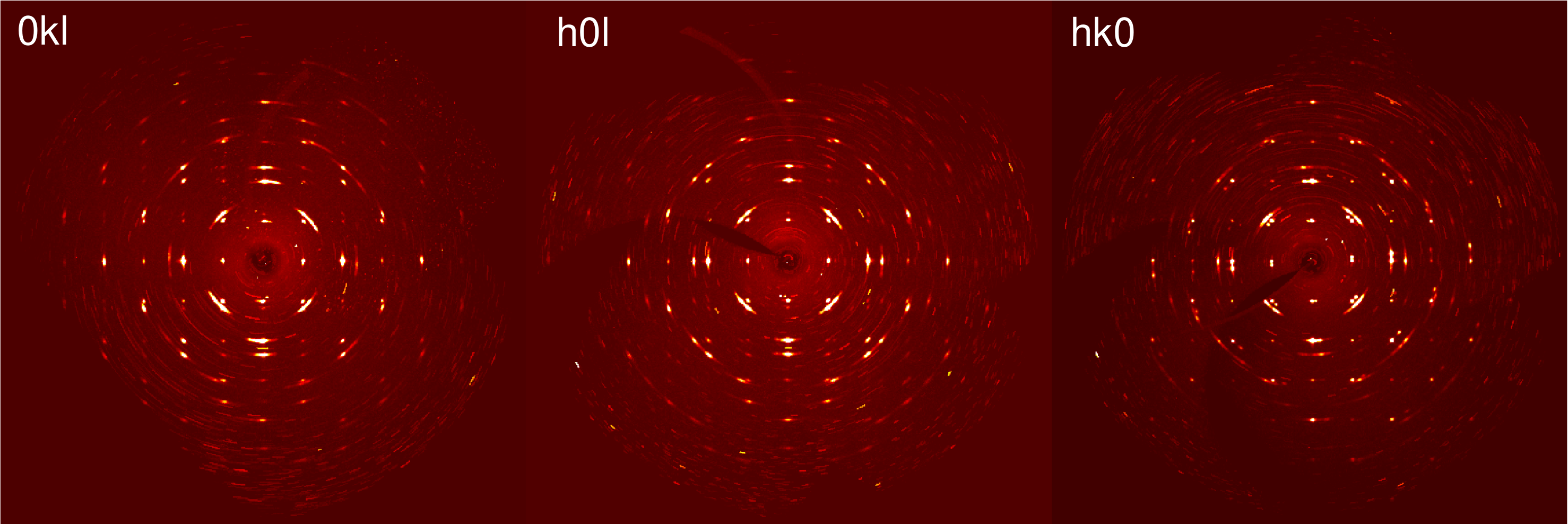}
\par\end{centering}
\caption{{\label{supp2}XRD maps  of the (0kl), (h0l) and (hk0) planes of the investigated infinite-layer  crystal.}}
\end{figure*}

As described in the materials and methods section, for the single crystal diffraction, a perovskite crystal was broken under high viscosity oil and a 20 \textmu m piece was mounted with grease on a loop made of Kapton foil (Micromounts, MiTeGen, Ithaca, NY). Diffraction data were collected with a SMART APEXI CCD X-ray diffractometer (Bruker AXS, Karlsruhe, Germany), using graphite-monochromated Mo-K$_{\alpha}$ radiation ($\lambda=0.71073\,$\AA) at room temperature $T=298(2)$ K. Figure~\ref{supp1} shows the XRD maps of the (hk0), (h0l) and (0kl) reciprocal lattice planes of the perovskite single crystal. In case of the infinite-layer nickelate, a 50 \textmu m piece was measured, which was broken off from the same polycrystal that was investigated with STEM. The obtained XRD maps of the reciprocal lattice planes are shown in Fig.~\ref{supp2}.

\section{Scanning electron microscopy analysis}

\begin{figure*}[h]
 \begin{centering}
\includegraphics[width=2\columnwidth]{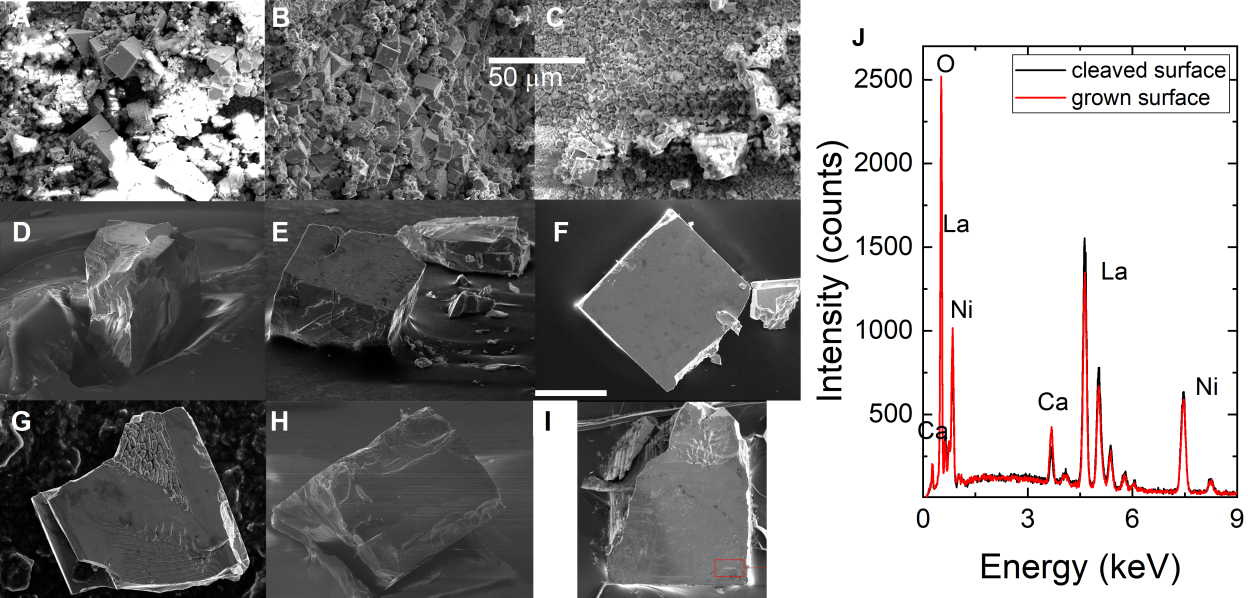} 
\par\end{centering}
\caption{{\label{SEM}SEM-SE images a-c) from the inside of the platinum ampule after the growth, d-h) selected perovskite single crystals, i) prevoskite crystal, where the highlighted red box depicts the cut STEM specimen and j) shows two selected EDS spectra of a perovskite crystal measured on an as-grown (red) and a cleaved surface (black), respectively.}}
\end{figure*}

The products of the growth and the Ca-substitution distribution inside the as-grown crystals were investigated through a comprehensive scanning electron microscope (SEM) analysis, with representative examples shown in Fig.~\ref{SEM}. While the single crystal refinements indicate a substitution level of 6\% and 8\% for two investigated crystals, energy-dispersive X-ray spectroscopy (EDS) reveals slightly higher values on the same investigated crystals with an average of 16(3)\% on as-grown surfaces and 10(5)\% on cleaved surfaces. Fig.~\ref{SEM}K displays a comparison of two EDS spectra measured on the same crystal on an as-grown and cleaved surface, respectively. Figure~\ref{SEM}J clearly shows reduced intensity of the Ca line and increased intensity on the La lines for the cleaved surface, resulting in a substitution reduction of 6(4)\% due to a concentration gradient.

\begin{figure*}[h]
 \begin{centering}
\includegraphics[width=1.8\columnwidth]{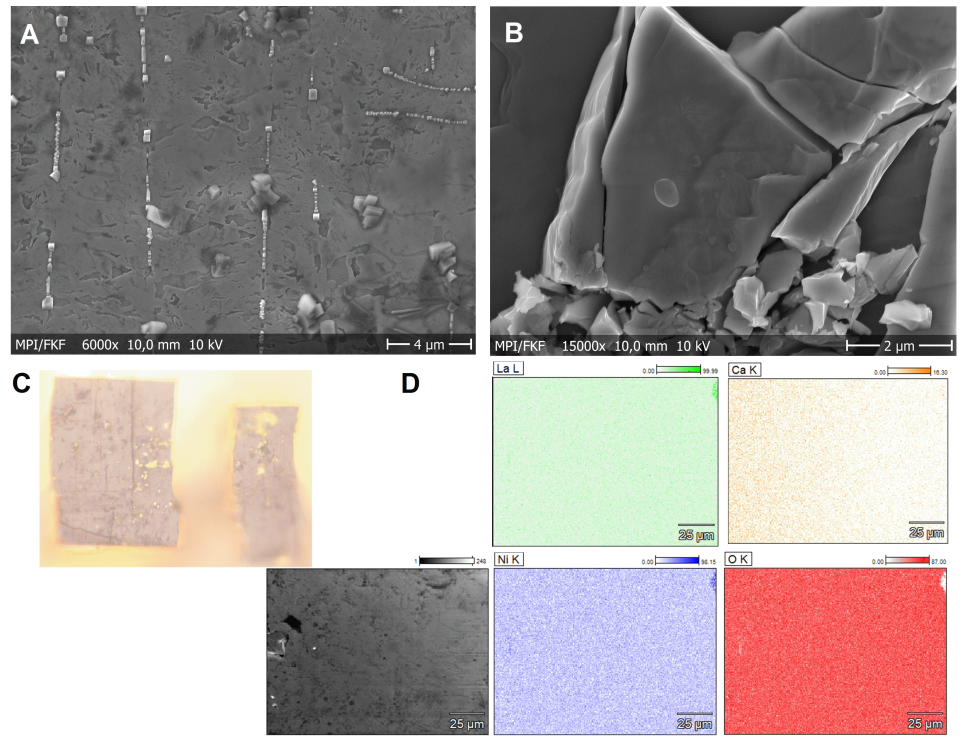} 
\par\end{centering}
\caption{{\label{EDS} a) SEM-BSE image of the surface of an as-grown pervoskite single-crystal. b) SEM-BSE image of cleavage planes of a pervoskite single-crystal. c) Optical microscope image showing the typical cleaving behavior observed for the perovskite crystals. d) SEM-SE image of a perovskite crystal (150 x 110 $\mu$m$^2$) and the corresponding elemental mapping of SEM-EDS spectra. The same color code as in the main text is used for the different elements.}}
\end{figure*}

On the surfaces of some as-grown crystals, we find CaO crystallized as tiny cubes along lines in backscattered electron (BSE) images (Fig.~\ref{EDS}A), as well as a NiO matrix showing the growth process on the surface of the perovskite crystal. On cleaved surfaces, however, we find no incorporation of CaO or NiO particles (Fig.~\ref{EDS}B). Instead, a highly accurate stoichiometric distribution is observed.

\section{Additional scanning transmission electron microscopy analysis}

\begin{figure*}[h]
 \begin{centering}
\includegraphics[width=1.5\columnwidth]{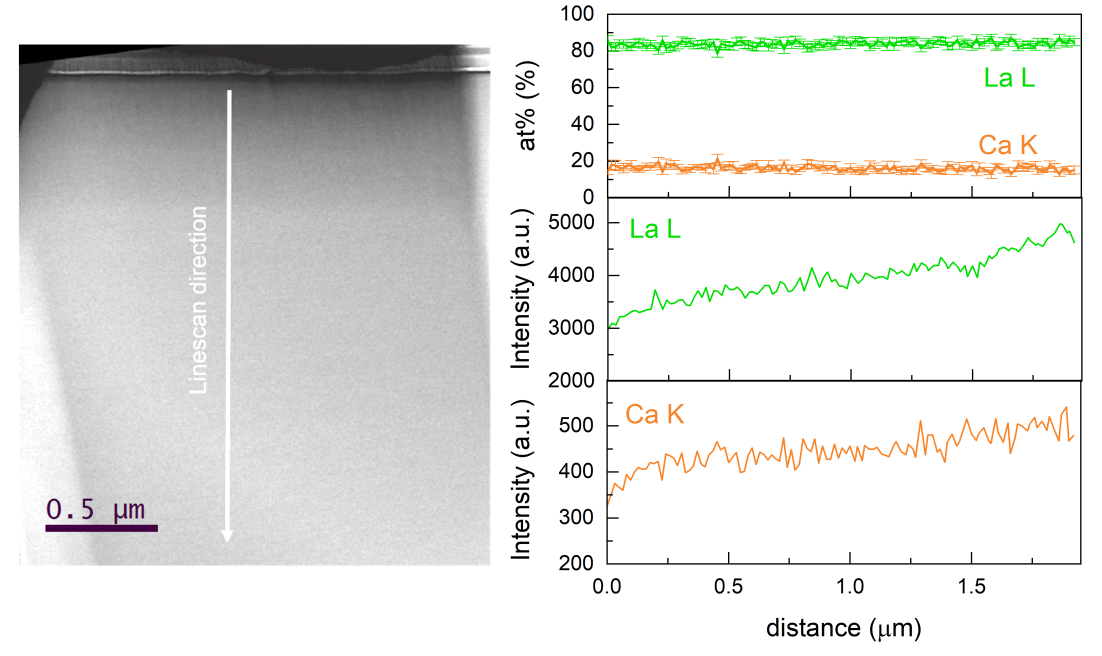} 
\par\end{centering}
\caption{{\label{domain}STEM-HAADF image of a large area of the prepared STEM-specimen from a perovskite crystal (left), investigated by STEM-EDS (La $L$-edge and Ca $K$-edge) with the lineprofiles of different elements shown on the right.}}
\end{figure*}

Performing scanning transmission electron microscopy (STEM) with EDS, we find a excellent agreement with the standard SEM-EDS analysis on the grown crystal surface, observing again a Ca substitution level of 16(2)\%. Fig.~\ref{domain} reveals that this substitution level is constant on a length scale of more than 2 \textmu m via a STEM-EDS linescan across the entire STEM-specimen, corresponding to a depth probe from the surface towards the center of the crystal. Note that the specimen is thinner near the surface and becomes thicker towards the center of the crystal, thus from the EDS-intensity profiles, both Ca and La signals get stronger as the specimen gets thicker away from the surface.

\end{document}